\documentstyle[epsfig,12pt]{article}
\begin{document}

\title{Mixed Symmetry Nuclear Shell Model}
\author{J. P. Draayer,$^{1}$ V. G. Gueorguiev\footnote{
On leave of absence from Institute of Nuclear Research and Nuclear Energy, Bulgarian Academy of Sciences, Sofia 1784, Bulgaria.},$^{1}$ Feng 
Pan,$^{1,2}$ and Yanan
Luo$^{1,2,3}$}
\date{\small \it $^{1}$Department of Physics and Astronomy, Louisiana 
State University, \\ Baton Rouge, LA 70803 \vspace{.1cm} \\ 
$^{2}$Department of Physics, Liaoning Normal University, \\ Dalian 
116029, P. R. China \vspace{.1cm} \\ $^{3}$Department of Physics, 
Nankai University, \\ Tianjin 300071, P. R. China}
\maketitle

\begin{abstract}
A mixed-symmetry nuclear shell-model scheme for carrying out calculations
in regimes where there is a competition between two or more modes is 
proposed. A
one-dimensional toy model is used to demonstrate the concept. The 
theory is then
applied to $^{24}Mg$ and $^{44}Ti$. For lower $pf$-shell nuclei such as
$^{44-48}Ti$ and $^{48}Cr$ there is strong $SU(3)$ symmetry breaking due to the
spin-orbit interaction. However, the quadrupole collectivity as 
measured by $B(E2)$
transition strengths in the yrast band remain high even though 
$SU(3)$ appears to
be broken. Some results for the so-called $X(5)$ symmetry that falls along the
$U(5) \leftrightarrow SU(3)$ leg of the Interacting Boson Model are also
considered. The results show that the mixed-symmetry concept is effective, even
when strong symmetry breaking occurs.

\end{abstract}

\section{The Mixed-Mode Concept}

Two dominant and often competing modes characterize the structure of
atomic nuclei. One is the single-particle structure that is demonstrated by the
validity of the mean-field concept; the other is the many-particle collective
behavior that is manifested through nuclear deformation. The 
spherical shell model
is the theory of choice whenever single-particle behavior dominates
\cite{Whitehead-shell model}. When deformation dominates, the Elliott 
$SU(3)$ model
or its pseudo-SU(3) extension is the natural choice \cite{Elliott}. 
This duality
manifests itself in two dominant components in the nuclear Hamiltonian:
respectively, the single-particle term, $H_{0}=\sum_{i}\varepsilon 
_{i}n_{i}$, and
a collective quadrupole-quadrupole interaction, $H_{QQ}=Q\cdot Q$. It 
follows that
a simplified Hamiltonian $H=\sum_{i}\varepsilon _{i}n_{i}-\chi Q\cdot 
Q$ has two
solvable limits.

\begin{figure}[htbp]
\centerline{\epsfxsize=4.0in\epsfbox{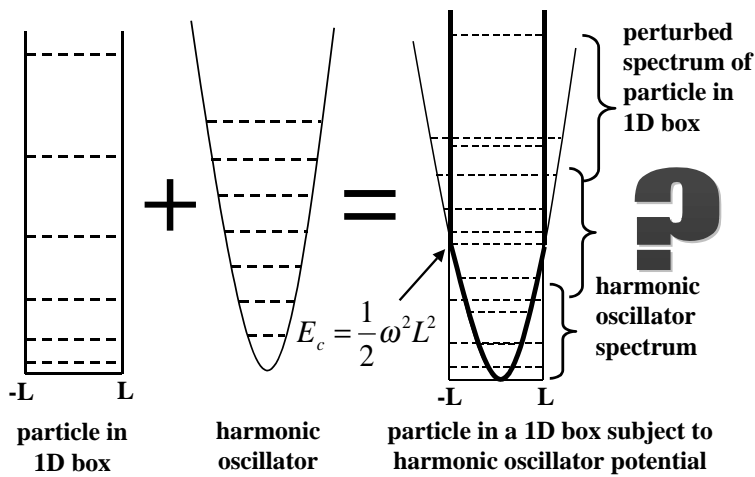}}
\caption{A schematic representation of the potential for a
particle in one-dimensional (1D) box at high energies and a
harmonic oscillator (HO) restoring force for low energies.}
\label{HO1Dbox potential}
\end{figure}

To probe the nature of such a system, one can consider a simpler
problem: the one-dimensional harmonic oscillator in a box of size 
$2L$  \cite{Vesselin's Ph.D. Dissertation}. As for real nuclei, this 
system has a finite volume and a restoring force of a harmonic 
oscillator type, $\omega ^{2}x^{2}/2$. For this model, shown in 
Figure \ref{HO1Dbox potential}, there is a well-defined energy scale 
that
measures the strength of the potential at the boundary of the box,
$E_{c}=\omega ^{2}L^{2}/2$. The value of $E_{c}$ determines the nature
of the low-energy excitations of the system. Specifically, depending on
the value of $E_{c}$ there are three spectral types:

\begin{itemize}
\item[(1)] For $\omega \rightarrow 0$ the spectrum is simply that of 
a particle in a box;
\item[(2)] At some value of $\omega$, the spectrum begins with 
$E_{c}$ followed by the spectrum of a particle in a box perturbed by 
the harmonic
oscillator potential;
\item[(3)] For sufficiently large $\omega $ the spectrum is that of a 
harmonic oscillator below $E_{c}$ which is followed by the perturbed 
spectrum of a particle in a box.
\end{itemize}

\begin{figure}[htbp]
\centerline{\epsfxsize=4.0in\epsfbox{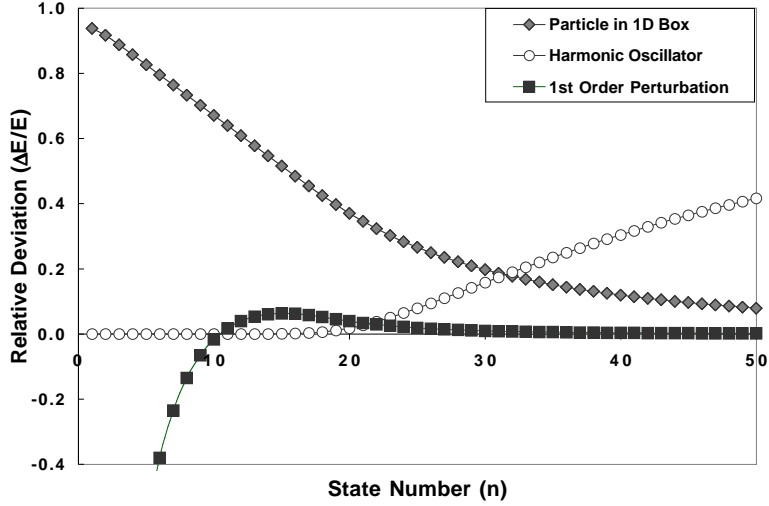}}
\caption{The relative deviations from the exact energy
eigenvalues for the 1D box plus HO potential
(Figure \ref{HO1Dbox potential}) with $\omega
=16$, $L=\pi /2$, $h/2\pi=m=1$. The open circles represent deviation of
the exact energy eigenvalue from the corresponding harmonic-oscillator
eigenvalue ($1-E_{ho}/E_{exact}$), the solid diamonds are the
corresponding relative deviation from the energy spectrum of a particle
in a 1D box, and the solid squares are the first-order perturbation
theory estimates using particle in a 1D box wavefunctions.}
\label{HO1Dbox spectrum}
\end{figure}

The last scenario (3) is the most interesting since it provides an example
of a two-mode system. For this case the use of two sets of basis
vectors, one representing each of the two limits, has physical appeal,
especially at energies near $E_{c}$. One basis set consists of the
harmonic oscillator states; the other set consists of basis states of a
particle in a 1D box. We call this combination a mixed-mode /
oblique-basis approach. In general, the oblique-basis vectors form a
nonorthogonal and overcomplete set. Even though a mixed spectrum is
expected around $E_{c}$, our numerical study that includes up to 50
harmonic oscillator states below $E_{c}$, shows that the first order
perturbation theory in energy using particle in a 1D box wave functions
as the zero order approximation to the exact functions works quite well
after the breakdown of the harmonic oscillator like spectrum. This
observation is demonstrated in Figure \ref{HO1Dbox spectrum} which
shows the relative deviations from the exact energy spectrum for a
particle in a 1D box.

\begin{figure}[htbp]
\centerline{\epsfxsize=4.0in\epsfbox{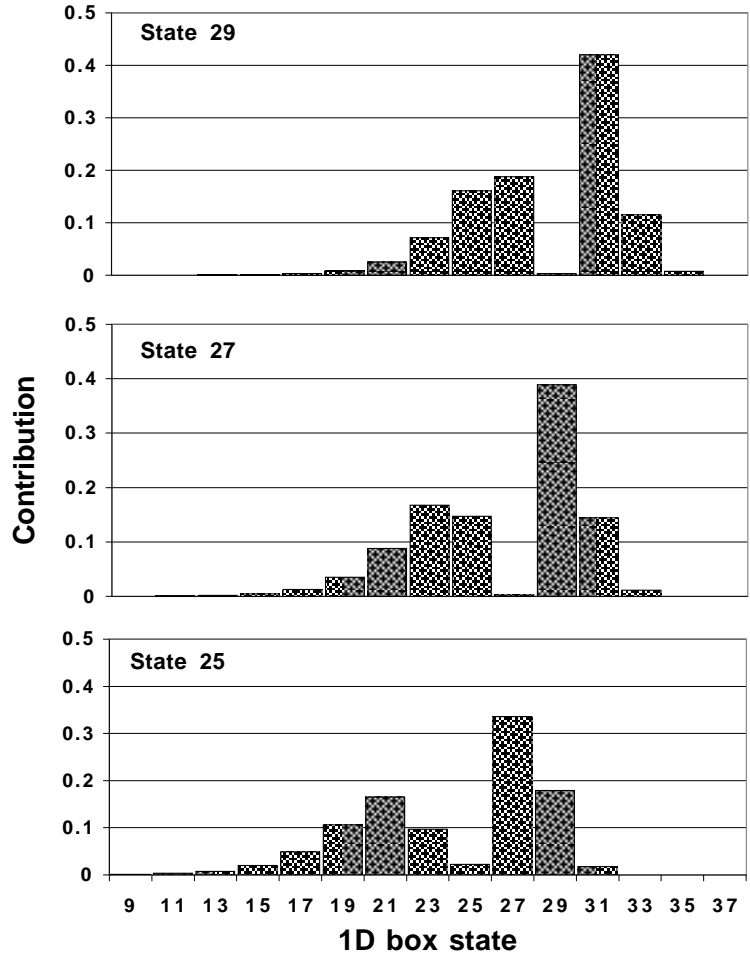}}
\caption{Bar charts that illustrate the similar, coherent structure of
the 25th, 27th and 29th exact eigenvectors in the basis of a free
particle in a 1D box. Parameters of the Hamiltonian are $\omega=16$,
$L=\pi/2$, $h/2\pi=m=1$.}
\label{Coherent mixing 1DHO}
\end{figure}

Although the spectrum seems to be well described using first order
perturbation theory based on a particle in a 1D box wave functions, the
exact wave functions near $E_{c}$ have an interesting structure. For
example, the zero order approximation to the wave function used to
calculate the energy may not be present at all in the structure of the
exact wave function as shown in Figure \ref{Coherent mixing 1DHO}.
Another feature also seen in Figure \ref{Coherent mixing 1DHO} is the
common shape of the distribution of the non-zero components along the
particle in a 1D box basis. The graph in Figure \ref{Coherent mixing
48Cr} shows this same effect in nuclei, which is usually attributed
to coherent mixing \cite{VGG SU(3)andLSinPF-ShellNuclei,Adiabatic
mixing}.

\section{Fermion-based Applications of the Theory}

\begin{figure}[htbp]
\centerline{\epsfxsize=4.5in\epsfbox{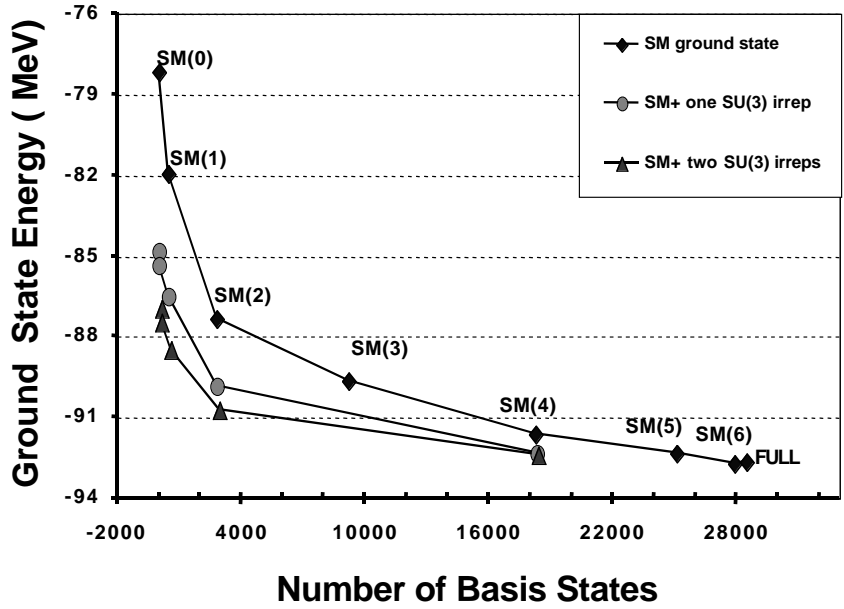}}
\caption{
The graph shows the calculated ground-state energy for
$^{24}Mg$ as a function of various model spaces. SM(n) denotes a
spherical shell model calculation with up to n particles outside of the
$d_{5/2}$ sub-shell. Note the dramatic increase in binding (3.3 MeV) in
going from SM(2) to SM(2)+(8,4)\&(9,2) (a 0.5\% increase in the
dimensionality of the model space). Enlarging the space from SM(2) to
SM(4) (a 54\% increase in the dimensionality of the model space) adds 4.2
MeV to the binding energy.}
\label{Mg24Oblique Con}
\end{figure}

\begin{figure}[htbp]
\centerline{\epsfxsize=4.5in\epsfbox{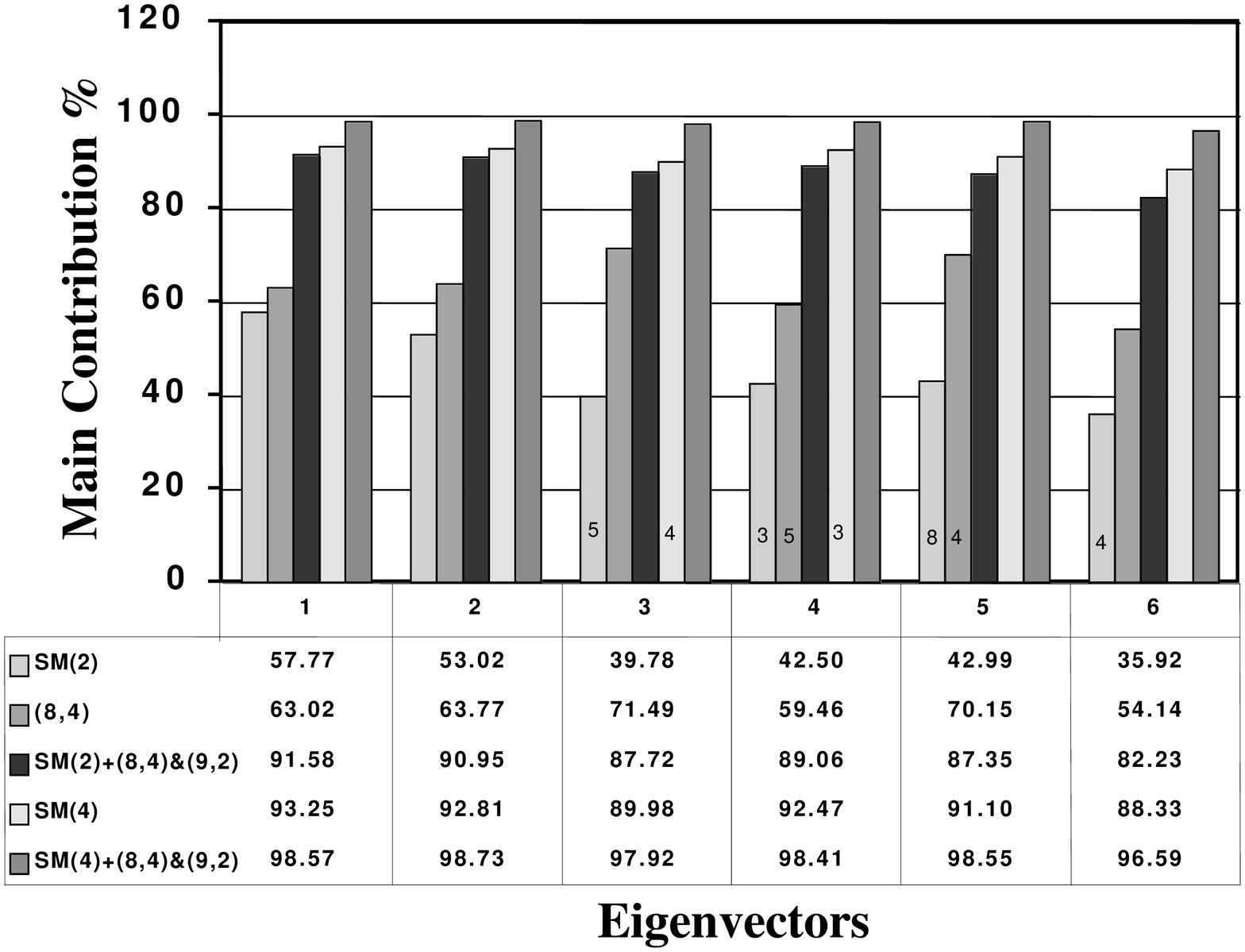}}
\caption{
Bar chart that shows representative overlaps of pure SM$(n)$, pure
SU(3), and oblique-basis results with the exact full $sd$-shell
eigenstates. A number within a bar denotes the state with the overlap
shown by the bar if it is different from the number for the exact
full-space calculation shown on the abscissa. For example, for SM(2)
the third eigenvector has the largest overlap with the fourth exact
eigenstate, not the third, while the fifth SM(2) eigenvector has
greatest overlap with the third exact eigenstate.}
\label{Mg24Oblique Bar}
\end{figure}

An application of the theory  to $^{24}Mg$
\cite{VGG 24MgObliqueCalculations}, using the realistic two-body interaction of
Wildenthal \cite{Wildenthal}, demonstrates the validity of the
mixed-mode concept. In this case the oblique-basis consists
of the traditional spherical states, that yields a diagonal
representation of the single-particle interaction, together with
collective $SU(3)$ configurations, that yields a diagonal representation
of the quadrupole-quadrupole interaction. The results shown in Figures
\ref{Mg24Oblique Con} and \ref{Mg24Oblique Bar} illustrate typical
outcomes. For example, a SM(2)+(8,4)\&(9,2) model space (third
bar in Figure \ref{Mg24Oblique Bar}) reproduces the binding energy
(within 2\% of the full-space result) as well as the low-energy
spectrum. For this case the calculated eigenstates have greater than
90\% overlap with the full-space results. In contrast, for a pure
$m$-scheme spherical shell-model calculation one needs about 60\% of
the full space, SM(4) -- the fourth bar in Figure \ref{Mg24Oblique Bar},
to obtain comparable results.

\begin{figure}[htbp]
\centerline{\epsfxsize=4.5in\epsfbox{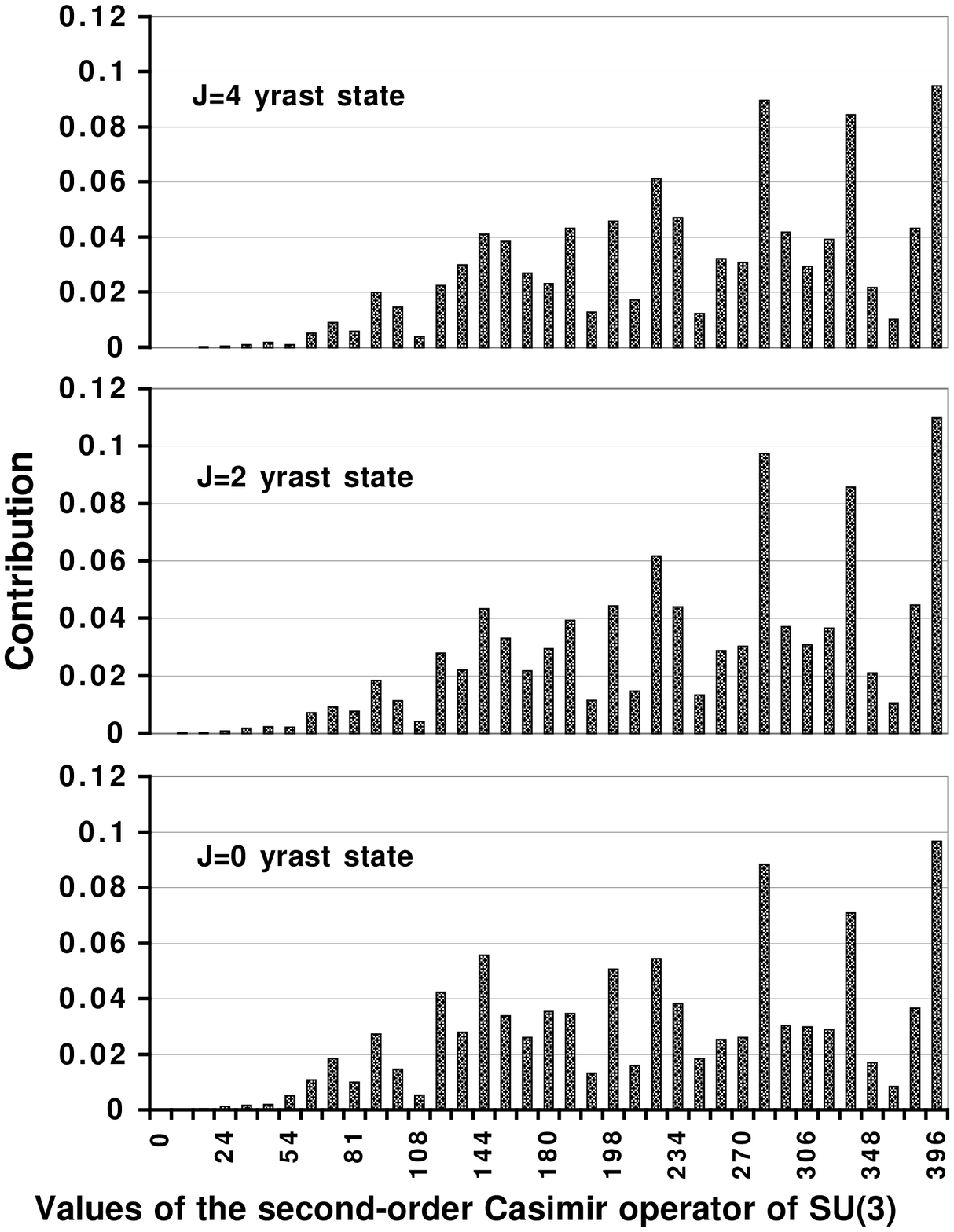}}
\caption{Bar charts that illustrate the very similar, coherent
structure of the first three yrast states in $^{48}Cr$ calculated using
the realistic Kuo-Brown-3 interaction ($KB3$). The horizontal axis is
$C_{2}$ of  $SU(3)$ while the height of each bar gives the contribution
of that configuration to the corresponding yrast state.}
\label{Coherent mixing 48Cr}
\end{figure}

Studies of the lower $pf$-shell nuclei $^{44-48}Ti$ and $^{48}Cr$ 
\cite{VGG SU(3)andLSinPF-ShellNuclei}, using the realistic 
Kuo-Brown-3 (KB3)
interaction \cite{KB3 interaction}, show strong $SU(3)$ symmetry breaking
due mainly to the single-particle spin-orbit splitting. Thus the KB3
Hamiltonian could also be considered a two-mode system. This is further
supported by the behavior of the yrast band B(E2) values that seems to
be insensitive to fragmentation of the $SU(3)$ symmetry. Specifically,
the quadrupole collectivity as measured by the B(E2) strengths remains
high even though the $SU(3)$ symmetry is rather badly broken. This has
been attributed to a quasi-SU(3) symmetry \cite{Adiabatic mixing} where
the observables behave like a pure $SU(3)$ symmetry while the true
eigenvectors exhibit a strong coherent structure with respect to each
of the two bases. This provides justification for further study
of the implications of mixed-mode shell-model studies.

Next we discuss oblique-basis type calculations for $^{44}Ti$ using the KB3
interaction \cite{KB3 interaction}. We confirm that the spherical shell
model (SSM) provides a significant part of the low-energy wave functions
within a relatively small number of SSMC while a pure $SU(3)$ 
shell-model with only few $SU(3)$ irreps is unsatisfactory. This is 
the
opposite of the situation in the lower $sd$-shell. Since the SSM yields
relatively good results for SM(2), combining the two basis sets yields
even better results with only a very small increase in the overall size of
the model space. In particular, results in a SM(2)+SU(3) model space
(47.7\% + 2.1\% of the full $pf$-shell space) are comparable with SM(3)
results (84\%). Therefore, as for the $sd$-shell, combining a few 
$SU(3)$ irreps with SM(2) configurations yields excellent results, 
such as correct
spectral structure, lower ground-state energy, and improved structure of
the wave functions. However, in the lower $sd$-shell $SU(3)$ is dominant and
SSM is recessive (but important) and in the lower $pf$-shell one finds the
opposite, that is, SSM is dominant and $SU(3)$ is recessive (but important).

For $^{44}Ti$ adding the $SU(3)$ to the SM(2) increases the model space
from 47.7\% to 49.8\% and gives results that are slightly better than the
SM(3) which is 84\% of the full space. In $^{24}Mg$ the position of the K=2
band head is correct for the SU(3)-type calculations but not for the
low-dimensional SM(n) calculations \cite{VGG 24MgObliqueCalculations}. In
$^{44}Ti$ it is the opposite, that is, the SM(n)-type calculations
reproduce the position of the K=2 band head while SU(3)-type calculations
cannot. Furthermore, the low-energy levels for the $SU(3)$ case are higher
than for the SM(n) case. Nonetheless, the spectral structure in the
oblique-basis calculation is good and the SM(2)+(12,0)\&(10,1) spectrum
($\approx$50\% of the full space) is comparable with the SM(3) result
(84\%).

The overlaps of SU(3)-type calculated eigenstates with the exact 
(full shell-model) results are, often less than 40\%, but the SM(n) 
overlaps are considerably bigger with SM(2)-type calculations 
yielding an 80\% overlap with the exact states while the results for 
SM(3) show overlaps greater than 97\%, which is consistent with the 
fact that SM(3) covers 84\% of the full space. On the other hand, 
SM(2)+(12,0)\&(10,1)-type calculations yield results that are as good 
as those for SM(3) in only about 50\% of the
full-space and SM(1)+(12,0)\&(10,1) overlaps are often bigger than the
SM(2) overlaps. The oblique-bases SM(2)+(12,0)\&(10,1) results for $^{44}Ti$
($\approx$50\%) yields results that are comparable with the SM(3) results
($\approx$84\%). In short, the oblique-basis scheme works well for
$^{44}Ti$, only in this case, in contrast with the previous results for
$^{24}Mg$ where $SU(3)$ was found to be dominant and SSM recessive, in the
lower $pf$-shell SSM is dominant and $SU(3)$ recessive.

\section{Boson-based Applications of the Theory}
In this section, we consider the transition from the $U(5)$ 
vibrational to the  $SU(3)$ rotational limit of the interacting boson 
model using a schematic
Hamiltonian. The behavior of low-lying energy levels and E2 transition rates
are studied in detail. The analysis shows that as one moves from the 
$SU(3)$ to the $U(5)$ limit the system transforms from one of maximum 
deformation to
one that is maximally triaxial and soft with the onset of strong triaxiality
occuring around the $X(5)$ critical point.

Understanding shape phase transitions of a finite many-body system is
paramount to understanding the system's underlying dynamics. The three
possible phases that can occur in the interacting boson model for nuclei
have been classified as $U(5)$, $SU(3)$, and $O(6)$ \cite{1B}.
The $U(5)\leftrightarrow SU(3)$ transitional description
of the rare-earth nuclei Nd, Sm, Gd, and Dy was first reported in \cite{5B},
including detailed results for most quantities of physical interest.
Evidence for coexisting phases at low energy in the spherical-deformed
transitional nucleus $^{152}$Sm was also analyzed using $U(5)\leftrightarrow
SU(3)$ transitional theory and the results show that the two phases coexist
in a very small region of parameter space around the critical point \cite{6B}.
Recently, since the discovery of the $X(5)$ symmetry in this region \cite{7B},
the spherical to axially deformed shape phase transition has attracted
further attention  \cite{8B}.

In order to take a close look at the $U(5)\leftrightarrow SU(3)$ shape phase
transition, we study transitional patterns of many physical 
quantities, such as low-lying energy levels, isomer shifts, E2 
transition rates, and some related quantities across the
$U(5)\leftrightarrow SU(3)$ leg of the Casten triangle. No attempt is made to
relate the results to realistic nuclei; rather, our purpose is to gain a
better understanding the nature of the $U(5)\leftrightarrow SU(3)$
transition. In the study, the schematic Hamiltonian
$$
\hat{H}=-c\left(x\hat{n}_{s}+{\frac{(1-x)}{{f(N)}}} \hat{Q}\cdot\hat{Q}
\right) \eqno(1)
$$
which has been suggested as a suitable form for describing nuclei in this
region is used, where the parameter $c>0$, $0\leq x\leq 1$ is the phase
parameter, $f(N)$ is a linear function of the total number of bosons $N$,
$\hat{n}_{s}=s^{\dagger}s$ is the number of $s$ bosons, and $\hat{Q}
=(s^{\dagger}\tilde{d}+d^{\dagger}\tilde{s})- {\frac{\sqrt{7}}{{2}}}
(d^{\dagger}\tilde{d})^{(2)}$. We note that $f(N)=1$ was used in 
\cite{6B}, while
$f(N)=4N$ was adopted in \cite{7B} and \cite{10B}. Also, the critical 
point $x_{{\rm c}}$
will be quite different for different choices of the function $f(N)$. The
Hamiltonian (1) is, up to a constant, equivalent to the one used in 
\cite{5B,6B,7B}
and \cite{10B} with the relation $\zeta=1-x$.

In order to diagonalize the Hamiltonian (1), we expand eigenstates of (1) in
terms of the $U(6)\supset SU(3)\supset SO(3)$ basis vectors $\vert
N(\lambda\mu)KL\rangle$ as

$$
\vert N L_{\xi}\rangle=\sum_{(\lambda\mu)K}C_{(\lambda\mu)K}^{L_{\xi}} \vert
N(\lambda\mu)KL\rangle,\eqno(2)
$$
where the quantum number $\xi$ indicating the $\xi$-th level with angular
momentum quantum number $L$ is introduced, and $C_{(\lambda\mu)K}^{L_{\xi}}$
is the expansion coefficients. Since the total number of bosons $N$ is fixed
for a given nucleus, the above eigenstates will also be denoted as
$\vert L_{\xi};x\rangle$ in the following, in which value of the phase
parameter $x$ is explicitly shown. In our calculation, the
orthonormalization process with respect to the band label $K$ and the phase
convention for the $SU(3)\supset SO(3)$ basis vectors proposed in 
\cite{11B,12B} is
used. By using analytic expressions for $U(6)\supset SU(3)$ reduced matrix
element of $s$-boson creation or annihilation operator \cite{11B} and an
algorithm  \cite{12B,13B} for generating the $SU(3)\supset SO(3)$ Wigner
coefficients, the eigen-equation that simultaneously determines the
eigenenergy and the corresponding set of the expansion coefficients
$\{C_{(\lambda\mu)K}^{L_{\xi}}\}$ can be established.

\begin{figure}[htbp]
\centerline{ \epsfig{file=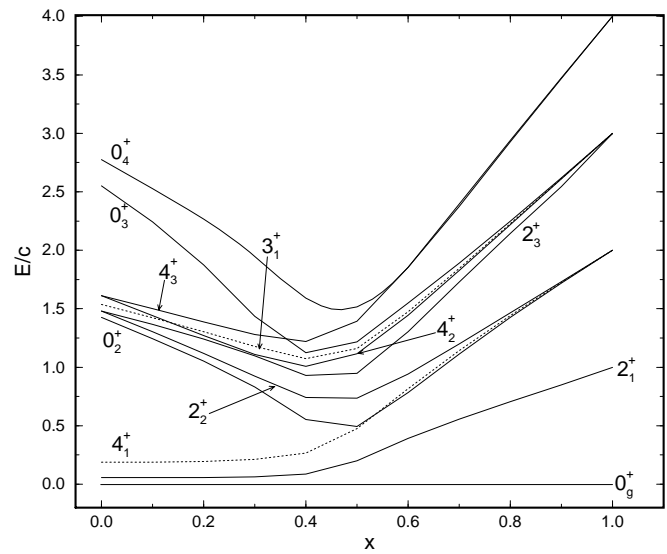, width=10.5cm} }
\caption{Some low-lying energy levels across the transitional region.}
\label{F1}
\end{figure}

To explore the transitional patterns, we fix the total number of bosons at
$N=10$ and allow the phase parameter $x$ to vary in the closed interval
$[0,1]$. The functional form of $f(N)$ is chosen to be the same as that used
in \cite{7B} with $f(N)=4N$ unless otherwise specifically noted. Some low-lying
energy levels as a function of $x$ are shown in Figure \ref{F1} from 
which one can
see that there is a minimum in the excitation energy around $x\sim 0.41-0.46$
, which corresponds to the spherical-deformed shape coexistence region which
is also referred to as the critical (phase transition) region. It can also
be seen that the minimum is not exactly the same for all the levels, but all
fall within the spherical-deformed shape coexistence region.

\begin{figure}[htbp]
\centerline{ \epsfig{file=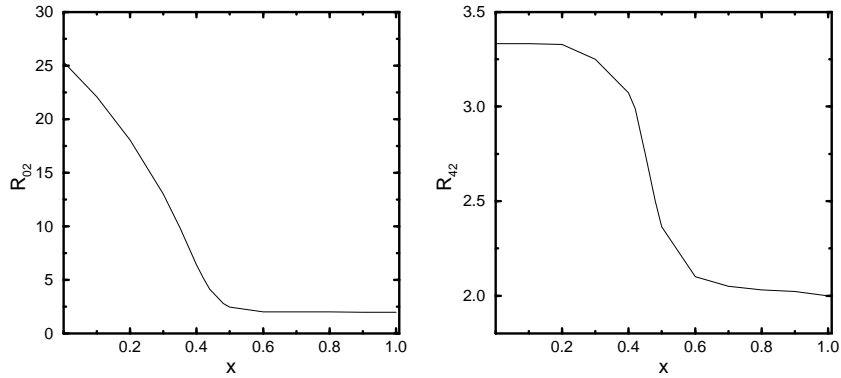,width=12cm} }
\caption{ Energy ratios $R_{02}$ and $R_{42}$ as functions of the
transitional parameter $x$.}
\label{F2}
\end{figure}

Two energy ratios, $R_{02}={E_{0^{+}_{2}}/E_{2^{+}_{1}}}$ and $R_{42}={
E_{4^{+}_{1}}/E_{2^{+}_{1}}}$, are shown as a function of $x$ in 
Figure \ref{F2}. The ratio $R_{02}$ drops rather precipitously from 
the axially deformed limit, $R_{02}=25.338$, to the spherical limit, 
$R_{02}\sim 2$, over the range  $0\leq x <~\sim 0.5$ and then remains 
at the spherical limit value for $x>0.5$. While the ratio $R_{42}$ 
drops rather smoothly as a function of $x$ from the deformed limit, 
$R_{42}=10/3$ when $0 \leq x <~\sim0.3$, to the
spherical limit, $R_{42}=2$ when $\sim0.6 < x \leq 1$. The sharpest change
occurs around the critical point $x_{{\rm c}}\sim 0.46$ when the absolute
value of the derivative of $R_{42}$ with respect to $x$ reaches the maximal
value.

\begin{figure}[htbp]
\centerline{ \epsfig{file=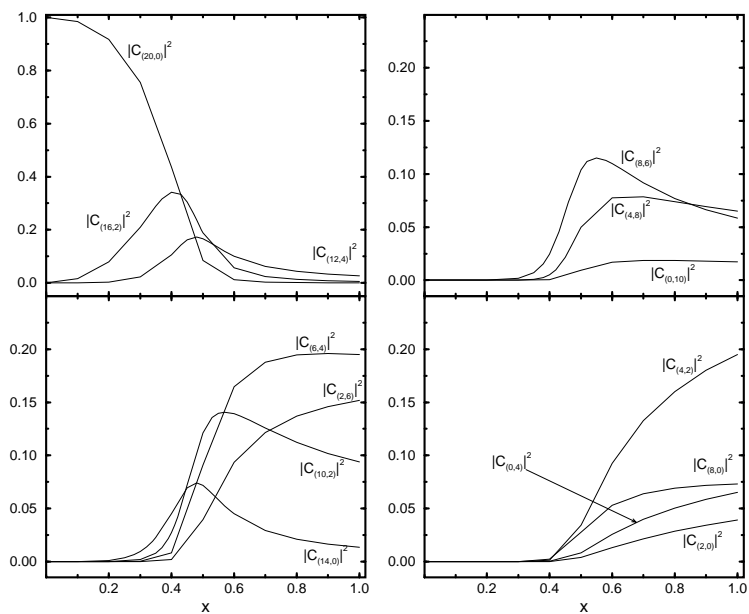,width=10.5cm} }
\caption{Amplitudes $\vert C_{(\lambda\mu)}\vert^{2}$ of the ground state as
functions of the transitional parameter $x$.}
\label{F3}
\end{figure}

To show how the transition occurs in the ground state, the amplitudes
$\vert C_{(\lambda\mu)}\vert^{2}$ are plotted as functions of $x$ in 
Figure \ref{F3}, which indicate that the most rapid changes in these 
amplitudes also occur within the coexistence region.

\begin{figure}[htbp]
\centerline{ \epsfig{file=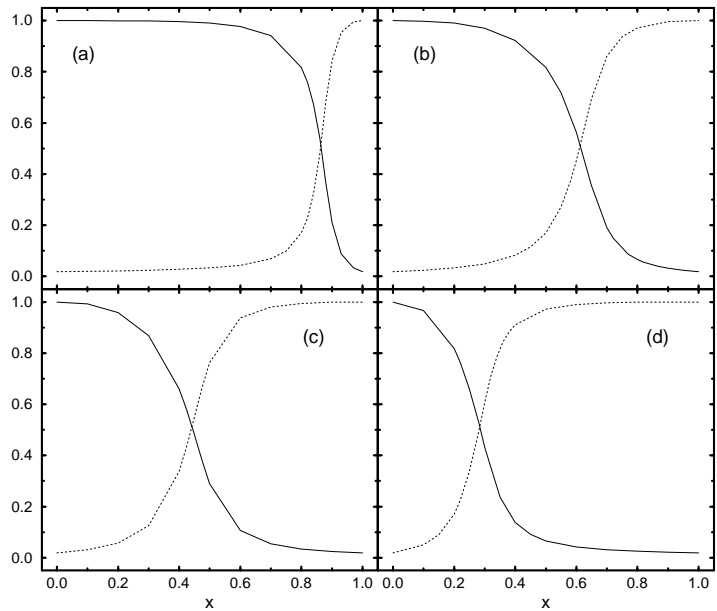, width=10.5cm} }
\caption{Overlaps of the ground state wavefunction, where the full line
shows the overlap $\vert\langle 0^{+}_{{\rm g}}; x=0\vert 0^{+}_{{\rm g}
};x\rangle\vert$, and the dotted line shows the overlap $\vert\langle 0^{+}_{
{\rm g}}; x=1\vert 0^{+}_{{\rm g}};x\rangle\vert$. (a) $f(N)=0.5N$; (b)
$f(N)=2N$; (c) $f(N)=4N$; (d) $f(N)=8N$. }
\label{F4}
\end{figure}

\begin{figure}[htbp]
\centerline{ \epsfig{file=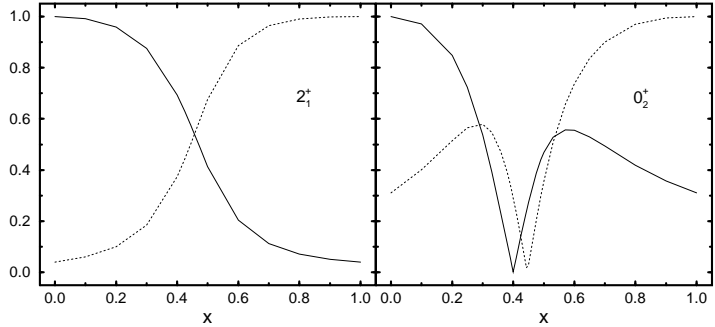, width=10.5cm} }
\caption{Typical overlaps of two excited states, where the full line shows
the overlap $\vert\langle L^{+}_{\xi}; x=0\vert L^{+}_{\xi};x\rangle\vert$,
and the dotted line shows the overlap $\vert\langle L^{+}_{\xi}; x=1\vert
L^{+}_{\xi};x\rangle\vert$.}
\label{F5}
\end{figure}

In order to explore the exact nature of the critical point with different
choices for the linear function $f(N)$, overlaps $\vert\langle 0^{+}_{{\rm g}
}; x\vert 0^{+}_{{\rm g}}; x_{0}\rangle\vert$ with $x_{0}=0$ or $1$ as
suggested in \cite{7B} with $f(N)=0.5N$, $2N$, $4N$, and $8N$, 
respectively, were calculated. The results are shown in Figure 
\ref{F4}, from which it can be clearly seen that the critical 
(crossing) point changes with different choices of
the function $f(N)$. The larger the $f(N)$ value, the smaller the critical
point value $x_{c}$. This conclusion confirms the early result shown 
in \cite{6B}, in which the critical point value is very large with 
$x_{c}=0.974$
corresponding to $f(N)=1$, while for $x_{c}\sim 0.46$ when $f(N)=4N$ with
$N=10$ is used \cite{7B,10B}. It should also be pointed out that the critical
point may differ from one excited state to the next for the same choice of
$f(N)$. Our calculation shows that the critical points for $0^{+}_{{\rm g}}$,
$2^{+}_{1}$, and $4^{+}_{1}$ are almost the same, but they are somewhat
different for $2_{3}^{+}$ and $3^{+}_{1}$. It should also be noted that
curves of the overlaps $\vert\langle L^{+}_{\xi}; x\vert
L^{+}_{\xi};x_{0}\rangle\vert$ with $x_{0}=0$ or $1$ even become irregular
for $0^{+}_{2}$, $2^{+}_{2}$, $4^{+}_{2}$, and $0^{+}_{3}$. Typical examples
of these curves for higher excited states are shown in Figure \ref{F5}, which
indicate that the critical point for higher excited states may be quite
different from that for ground state. This behavior will be seen in other
observables such as E2 transition rates since the critical behavior of the
initial and final states may quite different from that of ground state.

\begin{figure}[htbp]
\centerline{ \epsfig{file=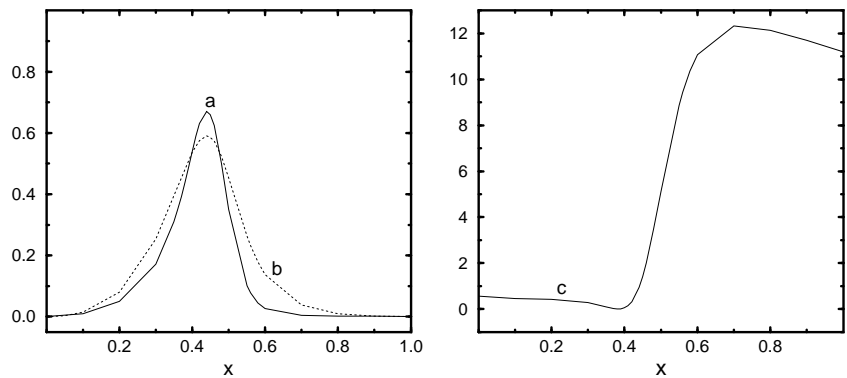, width=12cm} }
\caption{ Some B(E2)/$q^{2}_{2}$ values as functions of the transitional
parameter $x$, where curve a represents B(E2, $2^{+}_{3}\rightarrow 0^{+}_{
{\rm g}}$)/$q^{2}_{2}$, curve b represents B(E2, $2^{+}_{3}\rightarrow
2^{+}_{1}$)/$q^{2}_{2}$, and curve c represents B(E2, $2^{+}_{3}\rightarrow
0^{+}_{2}$)/$q^{2}_{2}$. }
\label{F7}
\end{figure}

\begin{figure}[htbp]
\centerline{ \epsfig{file=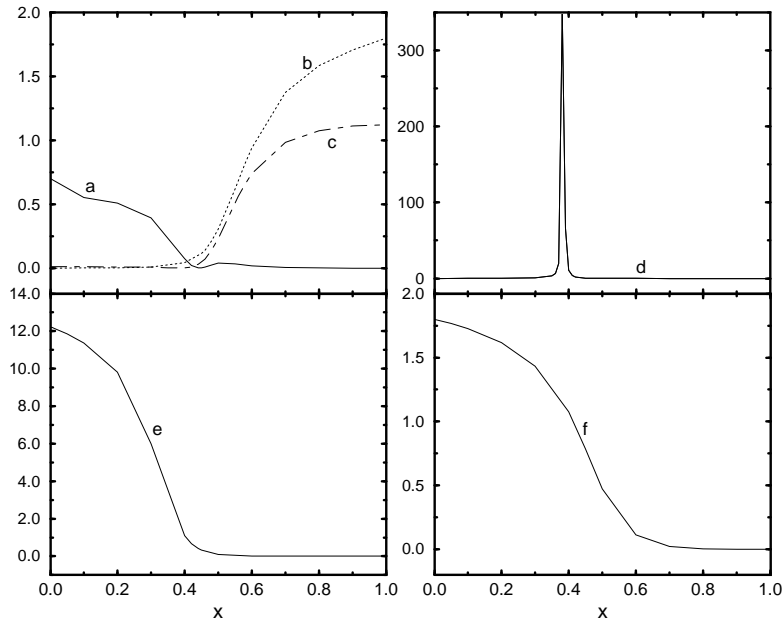, width=12cm} }
\caption{ Some B(E2) ratios as functions of the transitional parameter $x$,
where curve a represents B(E2; $2^{+}_{2}\rightarrow 0^{+}_{{\rm g}}$)/
B(E2; $2^{+}_{2}\rightarrow 2^{+}_{1}$), curve b represents B(E2;
$2^{+}_{2}\rightarrow 2^{+}_{1}$)/ B(E2; $2^{+}_{1}\rightarrow 0^{+}_{{\rm g}
} $), curve c represents B(E2; $2^{+}_{3}\rightarrow 0^{+}_{2}$)/ B(E2; $
2^{+}_{1}\rightarrow 0^{+}_{{\rm g}}$), curve d represents B(E2;
$2^{+}_{3}\rightarrow 0^{+}_{1}$)/ B(E2; $2^{+}_{3}\rightarrow 0^{+}_{2}$),
curve e represents B(E2; $2^{+}_{3}\rightarrow 2^{+}_{1}$)/ B(E2;
$2^{+}_{3}\rightarrow 0^{+}_{{\rm g}}$), and curve f represents B(E2; 
$3^{+}_{1}\rightarrow 2^{+}_{1}$)/ B(E2; $3^{+}_{1}\rightarrow 
4^{+}_{1}$).}
\label{F8}
\end{figure}

Various B(E2) values and ratios among the low-lying levels were 
studied using the E2 operator is $T(E2)=q_{2}\hat{Q}$, where $q_{2}$ 
is the effective charge. Some B(E2) values and ratios were found to 
be sensitive to the shape phase transition. Figure \ref{F7} provides 
three B(E2)
values, B(E2, $2^{+}_{3}\rightarrow 0^{+}_{{\rm g}}$), B(E2, 
$2^{+}_{3}\rightarrow 2^{+}_{1}$), and B(E2, $2^{+}_{3}\rightarrow 
0^{+}_{2}$
). There is a small peak around $x\sim 0.44$ for B(E2, $2^{+}_{3}\rightarrow
0^{+}_{{\rm g}}$) and B(E2, $2^{+}_{3}\rightarrow 2^{+}_{1}$), while there
is a saddle point around $x\sim 0.38$ for B(E2, $2^{+}_{3}\rightarrow
0^{+}_{2}$). As seen previously, the peak and saddle points in these B(E2)
values are different from the critical point of the ground state due to
different critical behavior of the excited states. Figure \ref{F8} 
provides six B(E2)
ratios, B(E2; $2^{+}_{2}\rightarrow 0^{+}_{{\rm g}}$)/ B(E2; 
$2^{+}_{2}\rightarrow 2^{+}_{1}$), B(E2; $2^{+}_{2}\rightarrow 
2^{+}_{1}$)/
B(E2; $2^{+}_{1}\rightarrow 0^{+}_{{\rm g}}$), B(E2; $2^{+}_{3}\rightarrow
0^{+}_{2}$)/ B(E2; $2^{+}_{1}\rightarrow 0^{+}_{{\rm g}}$), B(E2; 
$2^{+}_{3}\rightarrow 0^{+}_{1}$)/ B(E2; $2^{+}_{3}\rightarrow 
0^{+}_{2}$),
B(E2; $2^{+}_{3}\rightarrow 2^{+}_{1}$)/ B(E2; $2^{+}_{3}\rightarrow 
0^{+}_{{\rm g}}$), and B(E2; $3^{+}_{1}\rightarrow 2^{+}_{1}$)/ B(E2; 
$3^{+}_{1}\rightarrow 4^{+}_{1}$). These ratios all undergo noticeable
changes within the coexistence region. The most distinctive signature is
shown in B(E2; $2^{+}_{3}\rightarrow 0^{+}_{1}$)/ B(E2; $2^{+}_{3}
\rightarrow 0^{+}_{2}$), in which there is a giant peak around $x\sim 0.38$.

In order to study shape of nuclei around the critical point, we use the
relation between the Bohr variables $(\beta,\gamma)$ of the collective model
and the $(\lambda,\mu)$ labels that define the irreducible representation of
the $SU(3)$ \cite{14B}. In this algebraic approach, the Bohr variables
can be expressed as a functional of  $SU(3)$ Casimir operator of the second
order $\hat{\beta}=\beta_{0}\sqrt{\hat{C}_{2}(SU(3))+3}$, and
$\hat{\gamma}={\rm tan}^{-1}\left({\frac{\sqrt{3}(\hat{\mu}+1) }{{2\hat{
\lambda}+\hat{\mu}+3}}}\right)$, where $\beta_{0}$ can be taken as a 
constant and $\hat{\lambda}$ and $\hat{\mu}$ should be regarded as 
operators, of
which the results are usual $\lambda$ and $\mu$ values when they are applied
onto the basis vector of $SU(3)$.

\begin{figure}[htbp]
\centerline{ \epsfig{file=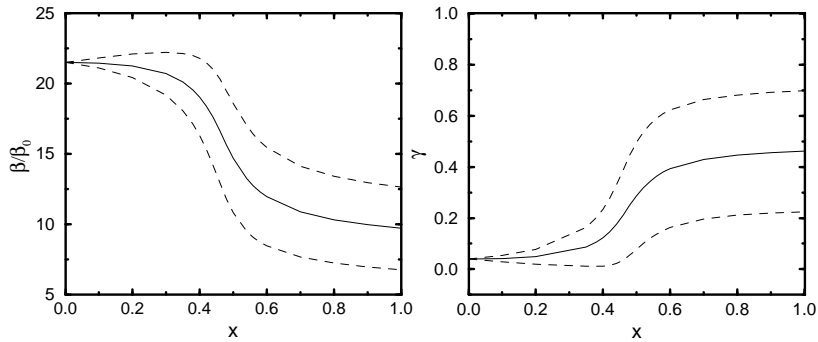, width=12cm} }
\caption{ The ground state expectation values of Bohr variables $\bar{\beta}$
and $\bar{\gamma}$ and the corresponding root mean square deviations
$\Delta(\beta)$ and $\Delta(\gamma)$, where the full line indicates the
expectation value $\bar{\beta}/\beta_{0}$ or $\bar{\gamma}$, while the
dotted lines show the corresponding root mean square deviations
$\pm\Delta(\beta)$ or $\pm\Delta(\gamma)$. }
\label{F9}
\end{figure}

Using these definitions for $\beta$ and $\gamma$, we can calculate 
expectation values, $\bar{\beta}=\langle 0^{+}_{{\rm g}};x\vert 
\hat{\beta}\vert 0^{+}_{{\rm g}};x\rangle$ and $\bar{\gamma}= \langle 
0^{+}_{{\rm g}};x\vert \hat{\gamma}\vert 0^{+}_{{\rm g}};x\rangle$ in 
the ground state, and the corresponding root mean square deviations 
$\Delta(\beta)$ and $\Delta(\gamma)$. These quantities can be used to 
display shape uncertainty as a function of $x$. It is obvious that 
$\Delta(\beta)$ and $\Delta(\gamma)$ are zero when nucleus is 
spherically
deformed with $x=0$ indicating there is a definite shape, while the shape
becomes less well defined as $x$ moves away from zero. The values of 
$\Delta(\beta)$ and $\Delta(\gamma)$ are not small in the spherical 
limit ($x=1$) due to the quadrupole vibration being not negligible. 
It can be seen from Figure \ref{F9} that there are also obvious 
changes in $\bar{\beta}$ and $\bar{\gamma}$ in the critical region 
around $x\sim 0.46$. Over the whole range, the magnitude of 
$\bar{\gamma}$ change is small. In contrast, $\Delta(\beta)$ reaches 
a maximum value around $x\sim 0.5$ which deviates a little from the 
critical point $x_{{\rm c}}\sim 0.46$ of the ground state, but is 
still near the critical region. This distinctive signature shows that 
nucleus is the softest in the critical region. This fact can help us 
to understand why there is a saddle region in most excited
levels. Since a nucleus in this region is comparatively soft, its shape can
be changed easily with very little energy. Hence, nuclei in this region can
easily be excited, which results in relatively smaller energy gaps in this
soft critical region.

In summary, transitional patterns from the vibrational, $U(5)$, to the
rotational, $SU(3)$, limit of the interacting boson model with a schematic
Hamiltonian have been studied. The transitional behavior of low-lying energy
levels, isomer shifts, E2 transition rates, and related quantities over the
whole $U(5)\leftrightarrow SU(3)$ transitional region were explored. The
results show that there are many distinctive signatures in the energy
levels, wavefunctions, isomer shift, BE2 values and ratios, and expectation
value of Bohr variables near the critical point. Generally speaking,
critical the behavior of excited states are different from that of ground
state, which may lead to different critical point for some physical
quantities that link the two. In comparison with the result shown in 
\cite{[9]}, a
nucleus with $X(5)$ symmetry can approximately be described by the 
$U(5)\leftrightarrow SU(3)$ transitional theory within the critical 
region.
Our analysis also shows that shapes of nucleus in the critical region is not
well-dfined; that is, a nucleus with $X(5)$ symmetry is soft.

\section{Conclusions}

Future research may provide justification for an extension of the
theory to multi-mode shell-model calculations. For example, an
immediate extension of the current scheme could use eigenvectors of the
pairing interaction \cite{Dukelsky et al-Pairing} within an $Sp(4)$
algebraic approach to the nuclear structure \cite{Sviratcheva-sp(4)} in
addition to collective  $SU(3)$ configurations and spherical shell model
states. Alternatively, Hamiltonian driven basis sets could be
considered. For example, the method could use eigenstates of near
closed shell nuclei obtained from a full shell-model calculation to
form Hamiltonian driven J-pair states for mid-shell
nuclei \cite{Heyde's-shell model}. This would mimic the Interacting
Boson Model (IBM) \cite{Iachello-1987} and the so-called broken-pair
theory \cite{Heyde's-shell model}. Likewise, the three exact limits of
the IBM \cite{MoshinskyBookOnHO} can be considered to comprise a
three-mode system. Nonetheless, the real benefit of the mixed-mode
approach is expected when the spaces encountered are too large to allow
for exact calculations.

\vskip .5cm This work was supported by the U.S. National Science Foundation
(0140300) and by the Natural Science Foundation of China (10175031).

\end{document}